\documentclass[letter,longauth,traditabstract]{aa} 
\usepackage{graphicx}
\usepackage{txfonts}
\newcommand{\GDRgra}{${\rm GDR}=65_{-18}^{+15}$}
\newcommand{\GDRac}{${\rm GDR}=287_{-42}^{+25}$}
\begin{document}
   \title{{\it HERschel} Inventory of The Agents of Galaxy Evolution (HERITAGE): the Large Magellanic Cloud dust \thanks{{\it Herschel} is an ESA space observatory with science instruments provided by European-led Principal Investigator consortia and with important participation from NASA.}}

   \titlerunning{I. HERITAGE: the LMC dust}

   \author{M. Meixner
    \inst{1, 2}
        \and    F. Galliano\inst{3}
    \and     S. Hony\inst{3}
    \and  J. Roman-Duval\inst{1}
     \and T. Robitaille\inst{4,5}
    \and    P. Panuzzo\inst{3}
    \and   M. Sauvage\inst{3}
    \and   K. Gordon\inst{1}
    \and   C. Engelbracht\inst{6}
     \and K. Misselt \inst{6}
    \and K. Okumura\inst{3}
   \and    T. Beck\inst{1}
    \and    J.-P. Bernard\inst{7}
    \and   A. Bolatto\inst{8}
    \and    C. Bot\inst{9}
    \and M. Boyer\inst{1}
   \and S. Bracker\inst{10}
   \and L.R. Carlson\inst{11}
    \and G. C. Clayton\inst{12}
   \and C.-H. R. Chen\inst{13}
    \and    E. Churchwell\inst{10}
    \and Y. Fukui\inst{14}
    \and M. Galametz\inst{3}
    \and J. L. Hora\inst{4}
    \and A. Hughes\inst{15}
  \and R.  Indebetouw\inst{13}
    \and F. P. Israel\inst{16}
    \and A. Kawamura\inst{14}
    \and F. Kemper\inst{17}
    \and S. Kim\inst{18}
    \and E. Kwon\inst{18}
   \and B. Lawton\inst{1}
    \and A. Li\inst{19}
   \and K. S. Long\inst{1}
    \and M. Marengo\inst{20}
    \and    S.C. Madden\inst{3}
  \and M. Matsuura\inst{21, 22}
    \and J. M. Oliveira\inst{23}
    \and T. Onishi\inst{24}
    \and M. Otsuka\inst{1}
    \and D. Paradis\inst{25}
    \and A. Poglitsch\inst{26}
   \and D. Riebel\inst{11}
    \and W. T. Reach\inst{25, 27}
    \and M. Rubio\inst{28}
    \and B. Sargent\inst{1}
    \and M. Sewi{\l}o\inst{1}
    \and J.D. Simon\inst{29}
    \and R. Skibba\inst{6}
   \and L.J. Smith\inst{1}
    \and S. Srinivasan\inst{30}
    \and   A.G.G.M. Tielens\inst{14}
   \and J. Th. van Loon\inst{23}
    \and B. Whitney\inst{31}
    \and P.~M.~Woods\inst{17}
              }

 \institute{Space Telescope Science Institute, 3700 San Martin Drive, Baltimore, MD 21218, USA \\
\email{meixner@stsci.edu}  
\and
Visiting Scientist at Smithsonian Astrophysical Observatory, Harvard-CfA, 60 Garden St., Cambridge, MA, 02138, USA 
\and CEA, Laboratoire AIM, Irfu/SAp, Orme des Merisiers, F-91191 Gif-sur-Yvette, France 
\and   Center for Astrophysics, 60 Garden St., MS 67, Harvard University, Cambridge, MA 02138, USA  
\and
Spitzer Fellow
\and Steward Observatory, University of Arizona, 933 North Cherry Ave., Tucson, AZ 85721, USA 
\and Centre d' \'{E}tude Spatiale des Rayonnements, CNRS, 9 av. du Colonel Roche, BP 4346, 31028 Toulouse, Fr  
\and  Department of Astronomy,  Lab for Millimeter-wave Astronomy, University of Maryland,  College Park, MD 20742-2421, USA 
\and  Observatoire Astronomique de Strasbourg, 11, rue de l'universite, 67000 STRASBOURG, France 
\and  Department of Astronomy, 475 North Charter St., University of Wisconsin, Madison, WI 53706, USA
\and  Johns Hopkins University, Department of Physics and Astronomy, Homewood Campus, Baltimore, MD 21218, USA 
\and Louisiana State University, Department of Physics \& Astronomy, 233-A Nicholson Hall, Tower Dr., Baton Rouge, LA 70803-4001, USA
\and Department of Astronomy, University of Virginia, and National Radio Astronomy Observatory,  PO Box 3818, Charlottesville, VA 22903, USA,  
\and  Department of Astrophysics, Nagoya University, Chikusa-ku, Nagoya 464-8602 , Japan 
\and Centre for Supercomputing and Astrophysics, Swinburne University of Technology, Hawthorn VIC 3122, Australia
\and Sterrewacht Leiden, Leiden University, P.O. Box 9513, NL-2300 RA Leiden, The Netherlands  
\and Jodrell Bank Centre for Astrophysics, Alan Turing Building, School of Physics \& Astronomy, University of Manchester, Oxford Road, Manchester M13 9PL, United Kingdom  
\and  Astronomy \& Space Science, Sejong University, 143-747, Seoul, South Korea   
\and 314 Physics Building, Department of Physics and Astronomy, University of Missouri, Columbia, MO 65211, USA
\and  Department of Physics and Astronomy, Iowa State University, Ames, IA, 50011, USA 
\and  Department of Physics and Astronomy, University College London, Gower Street, London WC1E 6BT, UK  
\and MSSL, University College London, Holmbury St. Mary, Dorking, Surrey RH5 6NT, UK
\and School of Physical \& Geographical Sciences, Lennard-Jones Laboratories, Keele University, Staffordshire ST5 5BG, UK
\and Department of Physical Science, Osaka Prefecture University, Gakuen 1-1, Sakai, Osaka 599-8531, Japan
\and Spitzer Science Center, California Institute of Technology, MS 220-6, Pasadena, CA  91125, USA
\and Max-Planck-Institut  f\"{u}r  extraterrestrische Physik, Giessenbachstra\ss e 85748 Garching,  Germany 
\and Stratospheric Observatory for Infrared Astronomy, Universities Space Research Association, Mail Stop 211-3, Moffett Field, CA 94035 
\and Departamento de Astronomia, Universidad de Chile, Casilla 36-D, Santiago, Chile
\and Observatories of the Carnegie Institution of Washington,   813 Santa Barbara St., Pasadena, CA, 91101 USA
\and   Astrophysique de Paris, Institute (IAP), CNRS UPR 341, 98bis, Boulevard Arago,  Paris, F-75014, Fr 
\and Space Science Institute,  4750 Walnut St. Suite 205, Boulder, CO 80301, USA
}

   \date{received: March 31, 2010; accepted: April 14, 2010 }

 
  \abstract{The {\it HERschel} Inventory of The Agents of Galaxy Evolution (HERITAGE) of the Magellanic Clouds will use dust emission to investigate the life cycle of matter in both the Large and Small Magellanic Clouds  (LMC and SMC). Using the {\it Herschel} Space Observatory's PACS and SPIRE photometry cameras, we imaged a $2^\circ \times 8^\circ$ strip through the LMC, at a position angle of $\sim$22.5$^\circ$  as part of the science demonstration phase of the {\it Herschel} mission. We present the data in all 5 {\it Herschel} bands:  PACS 100 and 160 $\mu$m and SPIRE 250, 350 and 500 $\mu$m.   We present two dust  models that both adequately fit the spectral energy distribution for the entire strip and both reveal that the SPIRE  500 $\mu$m emission is in excess of the models by  $\sim$6 to 17\%.  The SPIRE  emission follows the distribution of  the dust mass, which is derived from the model.  The  PAH-to-dust mass ($f_{PAH}$)  image of the strip reveals a possible enhancement in the LMC bar in agreement with previous work. We compare the gas mass distribution derived from the H{\small I} 21 cm and CO J=1-0 line emission maps to the dust mass map from the models  and derive gas-to-dust mass ratios (GDRs).   The dust model, which uses the standard graphite and silicate optical properties for Galactic dust,   has a very low \GDRgra  making it an unrealistic dust model for the LMC.  Our second dust model, which uses amorphous carbon instead of graphite,  has a flatter emissivity index in the submillimeter and  results in a \GDRac ~that is more consistent with a GDR inferred from extinction.}


   \maketitle
%

\section{Introduction}

The Large Magellanic Cloud (LMC) and
the Small Magellanic Cloud (SMC) are the best astrophysical
laboratories to study the lifecycle of the interstellar medium (ISM), because their proximity (50
kpc, e.g.\ Schaefer 2008; 61 kpc, Szewczyk et al.\  2009) permits detailed
studies of resolved  ISM clouds and their relation to  stellar
populations on  global scales, in an unambiguous manner, and as a
controlled function of environment.  Their sub-solar metallicities (${\rm Z}_{\rm LMC}\simeq 0.5\times {\rm Z}_\odot$, ${\rm Z}_{\rm SMC}\simeq 0.2\times{\rm Z}_\odot$; Dufour et al.\ 1982, Bernard et al.\ 2006) permit investigations on how processes  governing galaxy evolution depend on metallicity.  The {\it Herschel} Observatory  (Pilbratt et al.\ 2010) open-time key program, entitled {\it HERschel} Inventory of The Agents of Galaxy Evolution (HERITAGE) in the Magellanic Clouds, will  perform a uniform survey of    the LMC ($8^\circ \times 8.5^\circ$), SMC ($5^\circ \times 5^\circ$), and the Magellanic Bridge ($4^\circ \times 3^\circ$) with the Spectral and Photometric Imaging Receiver  (SPIRE) at 250, 350, and  500 $\mu$m  (Griffin et al.\ 2010) and with the Photodetector Array Camera and Spectrometer  (PACS) at 100 and 160 $\mu$m (Poglitsch et al.\ 2010).   The HERITAGE science goals are to study the life cycle of matter  in the Magellanic Clouds by probing the dust emission from the ISM and stars, which are the agents of galaxy evolution.  {\it Herschel} SPIRE and PACS images provide key insights into the life cycle of galaxies because the far-infrared and submillimeter emission from dust grains is an effective tracer of the  ISM dust, the most deeply embedded young stellar objects, and the dust ejected by evolved massive stars and supernovae. 
 
During the science demonstration phase (SDP), we imaged a long strip across the LMC that covers the entire range of interesting objects we expect to study with HERITAGE:  giant molecular  and  diffuse H$\,${\small I} clouds (Roman-Duval et al.\ 2010; Kim et al.\ 2010), H$\,${\small II}  regions (Hony et al.\ 2010),  young stellar objects (Sewi{\l}o et al.\ 2010), supernova remnants (Otsuka et al.\ 2010), and evolved stars (Boyer et al.\ 2010). In this paper, we  present the observing strategy, processing and resulting  data set (\S~2), the spectral energy distribution (SED) of the strip (\S~3) and the  global  spatial distribution of the dust and gas (\S~4).

\begin{figure*}
\includegraphics[width=17.0cm]{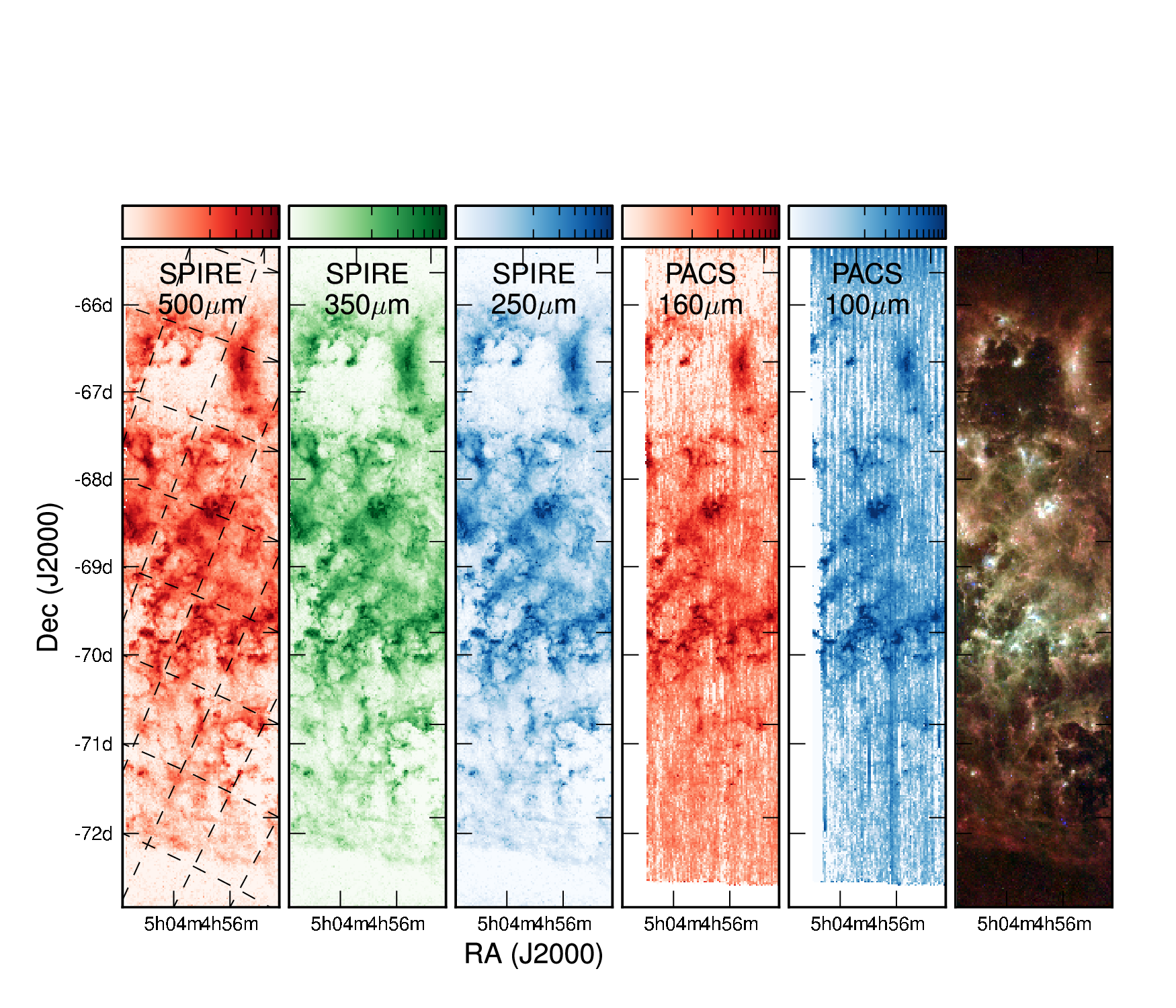}
\caption{The  HERITAGE LMC SDP strip in the five {\it Herschel} bands with color bars on top.  The strip has been reprojected to orient the figures up down, and represent the SPIRE coverage. The RA and Dec gridlines for all images here and in Fig. 3 are shown on the SPIRE 500 $\mu$m image. The PACS images are shifted because their coverage is shifted from the SPIRE data  in parallel mode. From left to right the color scale ranges and tick steps  for  the  arcsinh stretch (Lupton et al. 1999)  in units of MJy/sr:   0 to 31, steps of 5 (SPIRE 500 $\mu$m),  0 to 71, steps of 10  (SPIRE 350 $\mu$m), 0 to 161, steps of 20  (SPIRE 250 $\mu$m), 0 to 250, steps of 25 (PACS 160 $\mu$m) and -20 to  150, steps of 15 (PACS 100 $\mu$m).  Last on the right, a 3 color image with  SPIRE 350 $\mu$m in red, MIPS 70 $\mu$m in green and MIPS 24  $\mu$m in blue (Meixner et al.\ 2006).      }
\label{figdata}
\end{figure*}


\section{Observations  and data reduction}

We observed a $2^\circ \times 8^\circ$ strip through the LMC, at a position angle of $\sim$22.5$^\circ$ using the {\it Herschel}  Observatory   instruments SPIRE   and PACS in parallel observing mode. Observations began on November 22, 2009 at $\sim$23:00 h UT and lasted $\sim$18 hours.  Two 9 hour  astronomical observation requests (AORs) were constructed back-to-back in time to cover the region.   The observed wavelength bands  include PACS 100 and 160 $\mu$m and SPIRE  250, 350 and 500 $\mu$m (Fig.~\ref{figdata}).   

\subsection{PACS data reduction}

The PACS data  were reduced starting from the
level 0 product using the HIPE\footnote{HIPE is a joint development by the {\it Herschel} Science Ground
Segment Consortium, consisting of ESA, the NASA {\it Herschel} Science Center, and the HIFI, PACS and
SPIRE consortia.} version 2.0 data reduction software  (Ott 2010).   In
particular, bad and saturated pixels were masked, and a flat field correction and the photometric calibration were applied. 
The baseline signal level of each bolometer was estimated at the beginning and end
points of each scan leg which lie outside of the LMC.    For each pixel and each scan leg, a linear fit linking these two points was subtracted from the corresponding timeline.  The multi-resolution median transforms (MMT) deglitching algorithm was then applied  to correct sudden jumps in the timeline, or glitches, due to cosmic ray hits on the detector.

Initial two-dimensional maps were created for the 100  and 160 $\mu$m bands.  However, severe striping appeared along the scan direction of the PACS maps. To mitigate the stripes, we subtracted an image of the striping, produced by unambiguously identifying the power spectrum associated with the stripes, from the PACS images. The peak-to-peak variation was reduced from 0.021 to 0.011 in the PACS 100 and 160 $\mu$m images (Fig.~\ref{figdata}).  We do not use the PACS data for analysis in this paper, but anticipate that with cross-scans the PACS data will be viable for an extended map of the diffuse ISM.  Nevertheless, the de-striping was  effective for  small regions (Otsuka et al.\ 2010) and revealing point sources (Sewi{\l}o et al.\ 2010; Boyer et al.\ 2010).  

The  PACS 160 $\mu$m  brightness values are systematically 15\% lower than the corresponding MIPS 160 $\mu$m values from \cite{meixner2006}  using the adopted  HIPE 1.2 calibration tree which is consistent with the 20\% absolute flux calibration error for PACS  (Poglitsch et al. 2010). The accuracy of the PACS astrometry was estimated by comparing the positions of point sources detected in Spitzer   and PACS images of the LMC. We find offsets up to 6.5\arcsec, with no preferred direction, which have been traced to issues with the star trackers during very large scan maps.

\begin{figure}
\includegraphics[width=9.0cm]{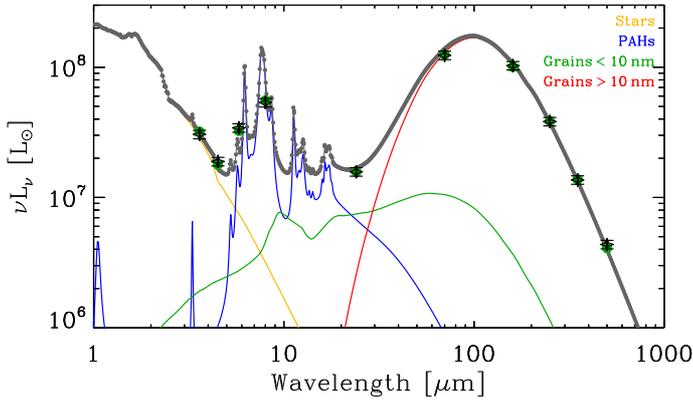}
\caption{ The spectral energy distribution of  the LMC HERITAGE SDP strip.  Open diamonds with error bars mark the fluxes from the SPIRE 500, 350, and 250 $\mu$m,  and the SAGE-LMC (Meixner et al. 2006) MIPS 160, 70 and 24 $\mu$m  and the IRAC 8, 5.8, 4.5 and 3.6 $\mu$m bands in solar luminosities.   The grey solid line  is the result of model 2 using the amorphous carbon dust  (see \S~3). The green filled points are the simulated photometry based on  model 2.  The  ISM dust emission components  are shown in blue, green and red according to the legend.  The yellow line is the stellar emission component. }
\label{figsed}
\end{figure}

\subsection{SPIRE data reduction}

The SPIRE data were processed  with HIPE version 2  (Ott  2010) and updated calibration products provided by the SPIRE team.    Using custom routines, we removed jumps in the timeline caused by co-occurring glitches that affect all detectors of a single array simultaneously and by jumps in the thermistor voltages.   We include the data at the end of each raster leg, during which the telescope turns around, in order to increase the coverage.  
For each scan, for each detector, we derived the median, baseline value of the measured flux at each end of the scan and subtract a linear baseline fitted to the two end points from the scans.   Images at 250, 350 and 500 $\mu$m  were constructed using the same reference point near the map center and are shown in Fig. \ref{figdata}.  In order to remove any instrumental residuals in the SPIRE images, we use regions at the ends of the map, assume they should have zero  emission and subtract a linear gradient from the images.   This process was done in a consistent way for all ancillary data (MIPS, IRAC images and HI 21 cm emission images) so that comparisons between the data sets during the analysis are done in a consistent fashion.

As recommended by the SPIRE ICC we multiplied the  maps by
the flux calibration correction factors   1.02, 1.05, 0.94 for 250, 350 and 500 $\mu$m, respectively.  The uncertainty in the final absolute point source  flux calibration is $\pm$15\%  (Swinyard et al.  2010).  The final SPIRE maps were converted from Jy per beam to MJy/sr using the effective beam areas of 9.28$\times 10^{-9}$,  1.74$\times 10^{-8}$, and  3.57$\times 10^{-8}$ sr  for  250, 350 and 500 $\mu$m.   
The positions of point sources in the 250 $\mu$m map were compared with their positions at 24 $\mu$m from MIPS revealing offsets up to 10\arcsec, similar to PACS. We corrected this problem with a small spherical rotation of the map, although some small residual astrometry errors remained, particularly far from the map centers.


\section{Spectral energy distribution of the strip}

The SED of the whole strip (Fig.~\ref{figsed}) is fit with a comprehensive dust  radiation  model described in detail by
\cite{galametz09}.   We adopt the Dale \&\ Helou (2002; Eq.~1) prescription, relating the
dust mass to the integrated energy density to which it is exposed in units of the solar neighborhood intensity ($2.2\times 10^{-5}{\rm W\,m^{-2}}$; Mathis et al.\ 1983). The standard model (model 1) we apply on the strip uses the
Galactic ISM dust size distribution and composition  of graphite and silicate grains from 
Zubko et al.\ (2004; BARE-GR-S model) who successfully fit the Galactic ISM dust emission out to submillimeter wavelengths.  
The dust emissivity index at the longest wavelengths is $\beta = 2$, which is also the assumed emissivity value for the SAGE-LMC ISM study by \cite{bernard2008}. 
However, as will be shown in Sect.~4, the grains of this model do not have 
enough emissivity in the submillimeter regime to fit the SPIRE fluxes, with 
a gas-to-dust mass ratio (GDR) consistent with the elemental abundances of the LMC.
As a consequence, we also created another model (model 2) replacing the 
graphites (of model 1) by amorphous carbons  from Rouleau \&\ Martin (1991; AC1).
This modified model has more emissivity at long wavelengths and therefore
gives a realistic GDR.  For both models, the PAH-to-dust mass fraction, $f_{PAH}$ is reported in units of the Galactic PAH fraction of $0.046$ (Draine \& Li 2007); i.e. $f_{PAH} = 1$ means Galactic PAH abundance. 

Both dust models provide an adequate fit to the SED of the strip.
For model 1, the PAH-to-dust mass fraction is slightly lower than the Galactic mass fraction ($f_{PAH}=0.85_{-0.06}^{+0.06}$), and its mass average stellar light intensity is twice the solar neighborhood intensity ($\langle U\rangle=2.0_{-0.5}^{+0.4}$).
Although H$\,${\sc ii} regions can show much hotter dust (Hony et al.\ 2010), it does not dominate the emission.  
We had to decrease the mass fraction of small grains ($<10$~nm) by a factor of 2, in order to fit the $24\;\mu$m flux of the diffuse ISM.
This has no significant effect on the total dust mass.
Model 2 gives a lower PAH mass fraction ($f_{PAH}=0.64_{-0.04}^{+0.07}$) and higher
starlight intensity ($\langle U\rangle= 9.1_{-1.3}^{+1.0}$).
Notably, the $500\;\mu$m flux is in excess compared to our SED  model of the whole strip by $\sim$17\%\ (model 1) and $\sim$6\%\ (model 2).
The origin of this excess is still unknown; however, such submillimeter excesses have been observed before in other magellanic irregular galaxies by e.g. 
\cite{galliano2005}.   
The spatial distribution and  potential location of this $500\;\mu$m flux excess is discussed in further details by \cite{gordon2010}.

\section{Distributions of dust and gas in the strip}

The emission at these  {\it Herschel} wavelengths (Fig.~\ref{figdata}) is dominated by dust emission from the ISM  and similar to the {\it Spitzer} MIPS 160 $\mu$m emission (see e.g.\ Bernard et al.\ 2008).  In fact,  the SPIRE maps have an identical appearance in the overall strip  which is expected because  they  trace the Rayleigh-Jeans tail of the ISM dust emission (Fig.~\ref{figsed}).  The three color  image in Fig.~\ref{figdata} shows the variation in color temperature of this ISM dust emission.   The SPIRE 350 $\mu$m (red) samples all dust emission, including some of the coldest dust ($\rm T\simeq10$~K), the MIPS 70 $\mu$m (green) is near  the SED peak,  and the  MIPS 24 $\mu$m (blue) represents the hottest  dust emission peaking in the H$\,${\small II}  regions.

\begin{figure}
\includegraphics[width= 9 cm]{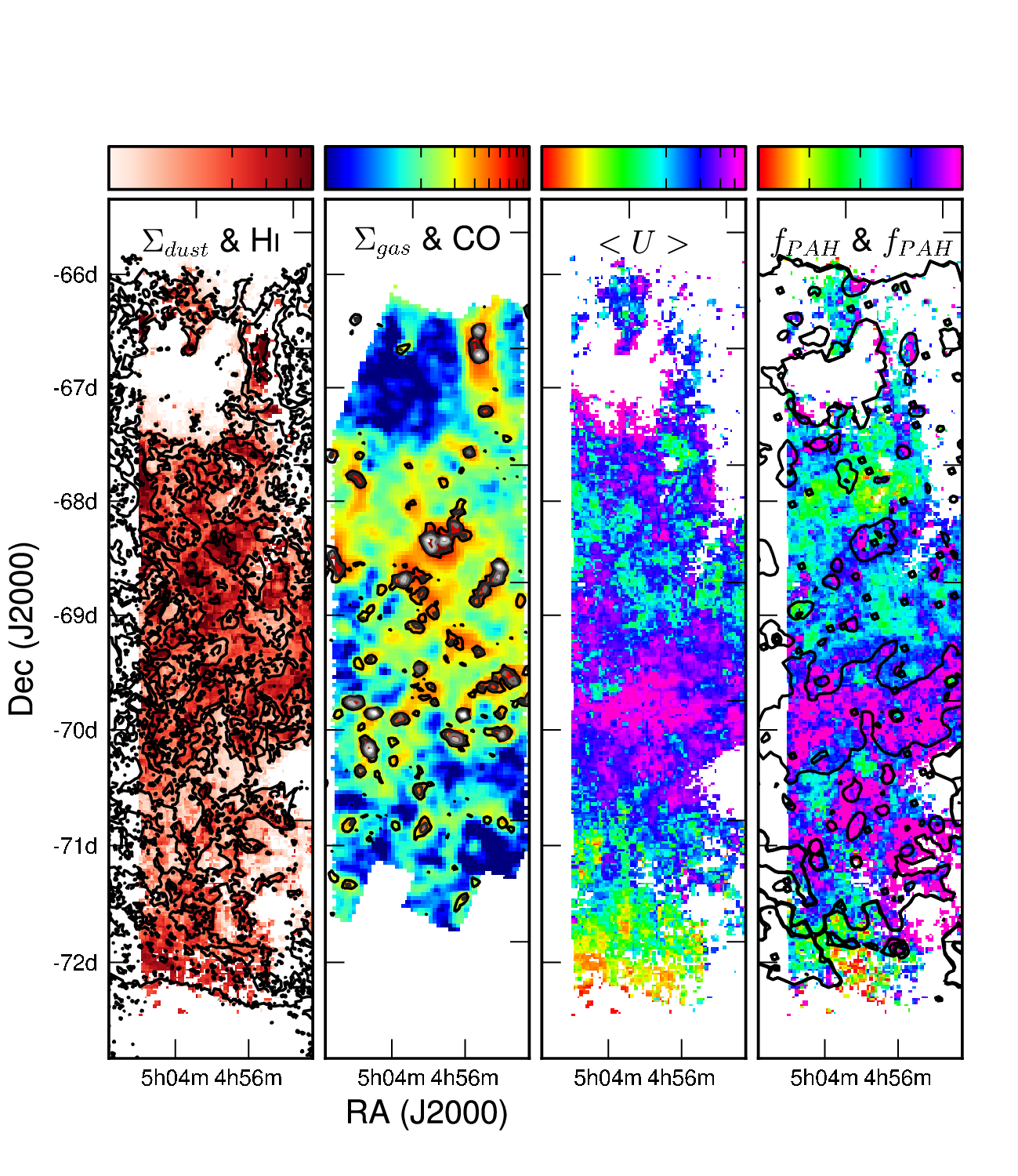}
\caption{Analysis images of the SDP strip based on model 2   (amorphous carbon, see \S~4) with color bars for each on top, described from left to right.    Dust surface mass, $\Sigma_{dust}$, in red  color arcsinh  scale ranging from   0 to 0.255, in tick steps of 0.05  M$_\odot$ pc$^{-2}$ with H{\small I} contours  at levels 1, 2 and 4 $\times 10^{21}$ H cm$^{-2}$ from \cite{kim2003}.  Gas surface mass, $\Sigma_{gas}$,  in jet color arcsinh scale  ranging from 0 to 90, with steps of 10 in  M$_\odot$ pc$^{-2}$ with CO contours at levels 0.8, 2, 4, 6 and 8 K kms/s  from \cite{fukui2008}. Average intensity  of the radiation field, $\langle U\rangle$,  in rainbow arcsinh color scale ranging from 0 to 20, with steps of 4 in units of the local solar neighborhood intensity.     Distribution of the PAH-to-dust mass fraction, $f_{PAH}$, in units of the Galactic fraction of 0.046,   as measured in the pixel based dust modeling with a  rainbow color linear  scale ranging from 0 to 0.8  in steps of 0.2 with the  comparable $f_{PAH}$ from \cite{paradis2009}  in black contours  levels of 0.007, 0.2 and 2. }
\label{figoverlay}
\end{figure}

The modelling described in Sect.~3 has also been applied to each pixel of the
strip (84\arcsec in angular size or 20~pc in linear size).  The  dust mass surface density, $\Sigma_{dust}$, average stellar light intensity, $\langle U\rangle$, and PAH-to-dust mass ratio, $f_{PAH}$,  derived from dust  model 2 are shown in Fig. \ref{figoverlay}.   The details of these models are described by Galliano et al. (in prep.).   We also derive a gas mass surface density ($\Sigma_{gas}$) image using the  H$\,${\small I} 21 cm emission from ATCA+Parkes by \cite{kim2003}, which traces the atomic neutral gas, and the CO J=1-0 emission from the NANTEN survey by \cite{fukui2008}, which traces the molecular gas.  The H{\small I} (1\arcmin\ resolution) and CO (2.6\arcmin\ resolution) maps were convolved to the IRIS 100 $\mu$m resolution (4.3\arcmin) using  Gaussian kernels of widths 4.18\arcmin\ and 3.42\arcmin\ respectively.  We used ${\rm X}_{CO} = 7\times 10^{20}\rm cm^{-2}K^{-1}km\,s^{-1}$  from Fukui et al.\ (2008) to derive the H$_2$ mass from CO.  We combine the H$\,${\small I}  and H$_2$  column densities  and adjust for the contribution of helium ($\times1.36$ more) to create the  total gas mass map in Fig.~3.
 
The dust mass distribution is spatially coincident with the SPIRE emission (Fig.~\ref{figoverlay}) and is discussed in detail by \cite{gordon2010}. The H$\,${\small I} 21 cm emission appears almost identical to the   dust mass image (Fig.~\ref{figoverlay}) and the similarity is quantified in more detail by \cite{kim2010}.    The CO$\,$J=$1-0$ emission coincides with the peaks in the  gas mass  (Fig.~\ref{figoverlay}) and two molecular clouds are discussed in more detail by \cite{romanduval2010}.   The average starlight intensity heating the dust, $\langle U\rangle$,  approximately follows  the IRAC 3.6 $\mu$m image (e.g.\ Meixner et al.\ 2006)  being enhanced in the stellar bar, and in the giant HII region complexes, such as N44, which is just north of the stellar bar.   The spatial distribution of $f_{PAH}$  derived by the model appears to be enhanced toward the stellar bar relative to the rest of the ISM  (Fig.~3) in agreement with the findings of \cite{paradis2009}.   We calculate that this enhancement is equivalent to  $\lesssim2000\;\rm M_\odot$  (model 1) or $\lesssim300\;\rm M_\odot$  (model 2)  in PAH.
However, it could be the result of the increased hardness of the starlight along the bar, mistaken for an increased abundance.

Over the area in common between the gas and dust mass images, we calculate the total gas mass and divide by the total dust mass to derive a GDR. For model$\,$1,  we calculate  a  \GDRgra.  This value is less than the Galactic GDR  value of 157 from \cite{zubko2004} which would be unexpected given the LMC's lower metallicity.  In fact, the model 1 GDR  is  a factor of  3 lower than the SAGE-LMC ISM study by \cite{bernard2008}, who  adopted similar Galactic dust properties.  It is also too low by a factor of three compared to UV extinction measurements of \cite{gordon2003}.   Thus, the addition of the SPIRE data demands too much dust, if we assume standard graphite and silicate optical properties which have been successfully used for Galactic dust models.  For model 2,  we calculate a \GDRac,  which is much more consistent  with prior observations.  Essentially the required dust mass for model 2 is smaller  by a factor $\sim$3 because it requires
a larger fraction of the dust to be illuminated
by intenser radiation fields and therefore less dust mass to account for the observed SPIRE measurements.  
The fact that the pixel to pixel averaged intensity $\langle U\rangle=7.6_{-3.8}^{+3.7}$ for model 2 is higher than for model 1 
($\langle U\rangle=1.0_{-0.4}^{+0.9}$) makes model 2 more consistent with 
the expected intense radiation conditions within the LMC.
The amorphous carbon emissivity is flatter in the submillimeter ($\beta<2$) which is in  agreement with the independent approach taken by \cite{gordon2010} for this data set and with the independent  analysis of the TOPHAT and DIRBE measurements of the LMC by \cite{aguirre2003}.  

Although our results indicate that the standard graphite and silicate optical properties for Galactic dust are not appropriate for the LMC dust,  our suggested amorphous carbon  and silicate dust model is not necessarily unique or the best model. 
The analysis of dust models begun in this paper  will need to be revisited when the full HERITAGE data set is available.  In particular a full exploration of parameter space for dust properties including composition and size distribution can be investigated in terms of  ISM environment and metallicity.

\begin{acknowledgements}
    
    We acknowledge financial support from the NASA {\it Herschel} Science Center, JPL contracts \# 1381522 \& 1381650.  M.R.  is supported by FONDECYT No1080335 and FONDAP No15010003.  We thank the contributions and support from the European Space Agency (ESA),  the PACS and SPIRE teams,  the {\it Herschel} Science Center and the NASA {\it Herschel} Science Center  (esp. A. Barbar and K. Xu) and the PACS and SPIRE instrument control centers,  without which none of this work would be possible.

    \end{acknowledgements}

\end{document}